%% file: cphwcs.tex
\documentclass{llncs}

\usepackage[utf8]{inputenc}

\usepackage[bookmarks = true, colorlinks=true, linkcolor = black, citecolor = black, menucolor = black, urlcolor = black]{hyperref}
\usepackage{graphicx}
\usepackage{amssymb,amsmath}
\usepackage{color}
\usepackage{listings}
\usepackage{tikz}
\usetikzlibrary{arrows}
\usepackage{verbatim}
\usepackage{theorem}
\usepackage{xy}
\xyoption{all} %para los diagramas

\theorembodyfont{\rmfamily}
\newtheorem{Thm}{Theorem}

\newtheorem{Defn}[Thm]{Definition}

\input macro

\newcommand{\SSReflect}{{\sc SSReflect }}

\begin{document}

\title{Computing Persistent Homology within Coq/SSReflect\thanks{Partially supported by Ministerio de Educaci\'on y Ciencia, project MTM2009-13842-C02-01, and by the European Union's
7th Framework Programme under grant agreement nr. 243847 (ForMath).}}

\titlerunning{Computing Persistent Homology within Coq/SSReflect}  % abbreviated title (for running head)
\author{J\'onathan Heras\inst{1} \and Thierry Coquand\inst{2} \and Anders M\"ortberg\inst{2} \and Vincent Siles\inst{2}}
\authorrunning{J. Heras \and T. Coquand \and A. Mörtberg \and V. Siles} % abbreviated author list (for running head)
%\tocauthor{author1, author2}
%

\institute{Department of Computing, University of Dundee, UK
\and Department of Computer Science and Engineering, Chalmers University of Technology and University of Gothenburg, Sweden
\\ \email{ jonathanheras@computing.dundee.ac.uk, \{coquand,mortberg,siles\}@chalmers.se}}

\maketitle              % typeset the title of the contribution

\begin{abstract}

\emph{Persistent homology} is one of the most active branches of
\emph{Computational Algebraic Topology} with applications in several contexts
such as optical character recognition or analysis of point cloud data. In this
paper, we report on the formal development of certified programs to compute
\emph{persistent Betti numbers}, an instrumental tool of persistent homology,
using the \Coq~proof assistant together with the \ssr~extension. To this aim it
has been necessary to formalize the underlying mathematical theory of these
algorithms. This is another example showing that interactive theorem provers have
reached a point where they are mature enough to tackle the formalization of
nontrivial mathematical theories.


\end{abstract}

\section{Introduction}\label{sec:intro}
\input{introduction}

\section{Mathematical background}\label{sec:phdaa}
\input{mathematics}

\section{An abstract formalization using \Coq/\SSReflect}\label{sec:aaais}
\input{abstract_algorithm}

\section{An effective certified implementation}\label{sec:aeci}
\input{effective}

\section{Experimental results}\label{sec:er}
\input{experimental}

\section{Conclusions and further work}
\input{conclusions}

\bibliographystyle{abbrv}
\bibliography{cphwcs}

\end{document}

%% file: macro.tex
\newcommand{\Coq}{{\sc Coq}}
\newcommand{\ssr}{{\sc SSReflect}}
\newcommand{\C}[1]{\mbox{\lstinline`#1`}}

\definecolor{dkblue}{rgb}{0,0.1,0.5} 
\definecolor{lightblue}{rgb}{0,0.5,0.5} 
\definecolor{dkgreen}{rgb}{0,0.4,0} 
\definecolor{dk2green}{rgb}{0.4,0,0} 
\definecolor{dkviolet}{rgb}{0.6,0,0.8}
\definecolor{shadethmcolor}{rgb}{0.9, 0.9,1}

%% file: introduction.tex
\emph{Persistent homology} is a branch of Algebraic Topology which appeared simultaneously in three works
during the last five years of the 20th century, see~\cite{DE95,FL99,R99}. Since that time it has become one of the central
tools in the context of Computational Algebraic Topology and several applications and extensions have been developed.

In a nutshell, persistent homology is a technique which allows one to study the \emph{lifetime} of
\emph{topological attributes}; this can be really useful in different contexts such as point cloud data~\cite{barcode08},
optical character recognition~\cite{cubicalhomology}, sensor networks~\cite{SG07} and surface reconstruction
from noisy samples~\cite{DG06}.

In this work, the main notions about persistent homology have been formalized using the \Coq~proof assistant~\cite{Coq}
together with the \ssr~extension~\cite{SSReflect}. During such a process we have proved relevant theorems like the
\emph{Fundamental Lemma of Persistent Homology}~\cite[pp. 152]{EH10}.
%% , which needs to be interpreted correctly before being formally proved.
%% Anders: Isn't this always the case?
Moreover, we have implemented certified programs to compute \emph{persistent Betti numbers},
an instrumental tool in the context of persistent homology.

The rest of this paper is organized as follows. The next section is devoted to present
the mathematical notions and results which will be formalized using \Coq/\ssr~in
Section~\ref{sec:aaais}. The effective algorithms to compute persistent homology
and some experiments performed with such programs are introduced respectively in Section~\ref{sec:aeci}
and Section~\ref{sec:er}. The paper ends with a section of conclusions
and further work.

The interested reader can consult the original and complete source code which can be found at~\cite{HCMS12}.

%% file: mathematics.tex
In this section, we briefly provide the necessary mathematical background needed to understand
the paper. We mainly focus on definitions, some of them are well known notions of
\emph{Algebraic Topology}, see~\cite{MacLane}, and the rest comes from \emph{persistent
homology theory}~\cite{ELZ02,ZC05,EH10}. We start by presenting simplicial complexes, a combinatorial
object which can be understood as a generalization of graphs to higher dimensions.

\subsection{Simplicial complexes}\label{math:sc}

Let $V$ be an \emph{ordered set}, called the \emph{vertex set}. An \emph{(abstract) simplex} over
$V$ is any finite subset of $V$. An \emph{(abstract) $n$-simplex} over $V$ is a simplex over $V$
whose cardinality is equal to $n+1$. Given a simplex $\alpha$ over $V$, we call \emph{faces} to 
the subsets of $\alpha$.

\begin{Defn}
An \emph{(ordered abstract) simplicial complex} over $V$ is a set of simplices $\mathcal{K}$
over $V$ such that it is closed by taking faces (subsets), that is, if
$\alpha \in \mathcal{K}$ all the faces of $\alpha$ are in $\mathcal{K}$ too.

Let $\mathcal{K}$ be a simplicial complex, the set $S_n(\mathcal{K})$ of $n$-simplices
of $\mathcal{K}$ is the set made of the simplices of cardinality $n+1$.
\end{Defn}

\begin{example}
Let us consider $V=(0,1,2,3,4,5)$. The small simplicial complex drawn in Figure \ref{diabolo} is
mathematically defined as the object:

$$\mathcal{K}=\left\{
               \begin{array}{l}
                 (0),(1),(2),(3),(4),(5),(0,1),(0,2),(1,2),(2,3)\\
                 (3,4),(3,5),(4,5),(0,1,2)\\
               \end{array}
             \right\}.$$

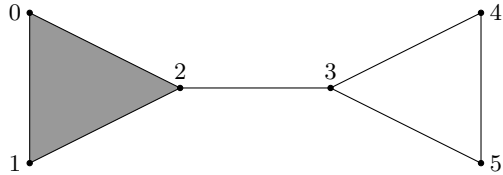
\begin{figure}
\centering
\begin{tikzpicture}
% The coordinates of each one of the points
\coordinate (1) at (0,0);
\coordinate (0) at (0,2);
\coordinate (2) at (2,1);
\coordinate (3) at (4,1);
\coordinate (5) at (6,0);
\coordinate (4) at (6,2);
% The unions among the points
\draw[fill=gray!80!white] (0)--(1)--(2)--cycle;
\draw (2)--(3);
\draw[fill=white] (3)--(4)--(5)--cycle;
% The points
\draw[fill=black] (0) circle (1pt and 1pt);
\draw[fill=black] (1) circle (1pt and 1pt);
\draw[fill=black] (2) circle (1pt and 1pt);
\draw[fill=black] (3) circle (1pt and 1pt);
\draw[fill=black] (4) circle (1pt and 1pt);
\draw[fill=black] (5) circle (1pt and 1pt);
% The names of the nodes
\draw (0) node[anchor=east] {0};
\draw (1) node[anchor=east] {1};
\draw (2) node[anchor=south] {2};
\draw (3) node[anchor=south] {3};
\draw (4) node[anchor=west] {4};
\draw (5) node[anchor=west] {5};
\end{tikzpicture}
\caption{Diabolo Complex}\label{diabolo}
\end{figure}

\end{example}

\begin{Defn}
Let $\mathcal{K}$ be a simplicial complex over $V$. Let $n$ and $i$ be two integers such that $n\geq 1$ and $0\leq i \leq n$.
Then the \emph{face operator} $\partial_i^n$ is the linear map $\partial_i^n:S_n(\mathcal{K}) \rightarrow S_{n-1}(\mathcal{K})$ defined by:
$$\partial_i^n((v_0,\ldots,v_n))= (v_0,\ldots,v_{i-1},v_{i+1}, \ldots ,v_n)$$
where the $i$-th vertex of the simplex is removed, so that a $(n-1)$-simplex is obtained.
\end{Defn}

\subsection{Chain complexes}\label{math:cc}

Now, we introduce a central notion in Algebraic Topology. Notions like rings, modules over a ring and module morphisms
(see~\cite{Basic-Algebra} for details) are assumed to be known.

\begin{Defn}\label{def:cc}
A \emph{chain complex} $C_\ast$ is a pair of sequences $(C_n,d_n)_{n\in\mathbb{Z}}$ where for every $n\in\mathbb{Z}$,
$C_n$ is a ${\mathcal R}$-module (with ${\mathcal R}$ a ring) and $d_n:C_n\rightarrow C_{n-1}$ is a module morphism, called the \emph{differential map},
such that the composition $d_nd_{n+1}$ is null (this is known as \emph{nilpotency} condition).

\noindent The module $C_n$ is called the module of \emph{$n$-chains}. The image
$B_n = im~d_{n+1} \subseteq C_n$ is the (sub)module of \emph{$n$-boundaries}. The kernel $Z_n = \ker d_n \subseteq C_n$ is the
(sub)module of \emph{$n$-cycles}.
\end{Defn}

Once we have defined the notions of simplicial complexes and chain complexes, we can define the link between them considering
$\mathbb{Z}_2$ as the ground ring. As $\mathbb{Z}_2$ is a field the chain groups are vector spaces.

\begin{Defn}
Let $\mathcal{K}$ be a simplicial complex over $V$. Then \emph{the chain complex $C_\ast(\mathcal{K})$ canonically associated with $\mathcal{K}$} is
defined as follows. The chain group $C_n(\mathcal{K})$ is the free $\mathbb{Z}_2$-module generated by the $n$-simplices of $\mathcal{K}$. In addition,
let $(v_0,\ldots,v_{n})$ be a $n$-simplex of $\mathcal{K}$, the differential of this simplex is defined as:
$$d_n := \sum\limits_{i=0}^n \partial_i^n.$$

$C_n(\mathcal{K})$ is a free module and the $n$-simplices form the \emph{standard basis} of it. Therefore, for all $n$ we can represent
$d_n:C_n(\mathcal{K}) \rightarrow C_{n-1}(\mathcal{K})$ relative to the standard basis of the chain groups as a $\mathbb{Z}_2$ matrix. Such
a matrix is called \emph{the n-th incidence matrix} of a simplicial complex.
\end{Defn}

Let us present an example in order to clarify the notion of chain complex canonically associated with a simplicial complex.

\begin{example}
Let $\mathcal{K}$ be the simplicial complex defined in Figure~\ref{diabolo}. The chain complex $C_\ast(\mathcal{K})$ canonically associated with $\mathcal{K}$ is:
$$\cdots \rightarrow 0 \rightarrow C_2(\mathcal{K}) \xrightarrow{d_2} C_1(\mathcal{K}) \xrightarrow{d_1} C_0(\mathcal{K}) \rightarrow 0 \rightarrow \cdots$$
\noindent where the $3$ associated chain groups are:
\begin{itemize}
  \item $C_0(\mathcal{K})$, the free $\mathbb{Z}_2$-module on the set of $0$-simplices (vertices)\\ $\{(0), (1), (2), (3), (4), (5)\}$.
  \item $C_1(\mathcal{K})$, the free $\mathbb{Z}_2$-module on the set of $1$-simplices (edges)\\ $\{(0, 1), (0, 2), (1, 2), (2, 3), (3, 4), (3, 5), (4, 5)\}$.
  \item $C_2(\mathcal{K})$, the free $\mathbb{Z}_2$-module on the set of $2$-simplices (triangles)\\ $\{(0, 1,2)\}$.
\end{itemize}

\noindent and the first incidence matrix, $d_1$, is:

$$
\bordermatrix{&(0,1) & (0,2) & (1,2) & (2,3) & (3,4) & (3,5) & (4,5)\cr
(0) & 1 & 1 & 0 & 0 & 0 & 0 & 0  \cr
(1) & 1 & 0 & 1 & 0 & 0 & 0 & 0 \cr
(2) & 0 & 1 & 1 & 1 & 0 & 0 & 0\cr
(3) & 0 & 0 & 0 & 1 & 1 & 1 & 0 \cr
(4) & 0 & 0 & 0 & 0 & 1 & 0 & 1 \cr
(5) & 0 & 0 & 0 & 0 & 0 & 1 & 1 \cr}
$$

\end{example}

Finally, we introduce one of the most important notions in the context of Computational Algebraic Topology.
Given a chain complex $C_\ast=(C_n,d_n)_{n \in \mathbb{Z}}$, the identities $d_{n-1}\circ d_n= 0$ mean that the inclusion
relations \(B_n \subseteq Z_n\) hold, that is,  every boundary is a cycle (the converse is generally not true). Thus the next definition
makes sense.

\begin{Defn}
The \emph{$n$-th homology group} of $C_\ast$, denoted by $H_n(C_\ast)$, is defined as the quotient $H_n(C_\ast)=Z_n/B_n$.
The elements of $H_n(C_\ast)$ are called \emph{$n$-dimensional homology classes} of $C_\ast$.

The \emph{$n$-th Betti number} of $C_\ast$, denoted by $\beta_n(C_\ast)$, is the \emph{rank} of the $n$-th homology group of
$C_\ast$.
\end{Defn}

In an intuitive sense, the $n$-th Betti number of an object $X$ measures the number of $n$ holes of $X$; to be more concrete,
$\beta_0$ measures the number of connected components, and the Betti numbers $\beta_n$, with $n>0$, measure higher dimensional
connectedness.

The homology groups of a simplicial complex $\mathcal{K}$ are the ones associated with the chain complex $C_\ast(\mathcal{K})$.
Moreover, Betti numbers of a simplicial complex can be easily computed considering the representation of
the differential maps as matrices using the formula:

\begin{equation}\label{betti_rank}
\beta_n(C_\ast(\mathcal{K})) = ns - rank (d_n) - rank (d_{n+1})
\end{equation}

\noindent where $ns$ is the number of $n$-simplices.

\subsection{Persistent Homology}\label{math:ph}

We end this section by introducing the instrumental notions in persistent homology theory. A more detailed description of this
theory can be found in~\cite{ELZ02,ZC05,EH10}.

\begin{Defn}
Let $\mathcal{K}$ be a simplicial complex, a \emph{subcomplex} of $\mathcal{K}$ is a subset $\mathcal{L} \subseteq \mathcal{K}$ that is also
a simplicial complex. A \emph{filtration} of $\mathcal{K}$ is a nested subsequence of complexes:
$$K^0 \subseteq K^1 \subseteq \ldots \subseteq K^m = \mathcal{K}$$
\end{Defn}

\begin{figure}
\centering
\begin{tikzpicture}

 \draw (0,0) node[rectangle,draw]{
\begin{tikzpicture}

\draw (0,-1) node {$K^0$};
\draw (4,-1) node {$K^1$};
\draw (8,-1) node {$K^2$};
\draw (0,-3) node {$K^3$};
\draw (4,-3) node {$K^4$};
\draw (8,-3) node {$K^5 = {\cal K}$};
%1
\draw (0,0) node[rectangle,draw ]{
\begin{tikzpicture}[scale=0.4]
\coordinate (1) at (0,0);
\coordinate (0) at (0,2);
\coordinate (2) at (2,1);
\coordinate (3) at (4,1);
\coordinate (5) at (6,0);
\coordinate (4) at (6,2);
% The points
\draw[fill=black] (0) circle (1pt and 1pt);
\draw[fill=black] (1) circle (1pt and 1pt);
\draw[fill=black] (2) circle (1pt and 1pt);
\draw[fill=white,color=white] (3) circle (1pt and 1pt);
\draw[fill=white,color=white] (4) circle (1pt and 1pt);
\draw[fill=white,color=white] (5) circle (1pt and 1pt);
% The names of the nodes
\draw (0) node[anchor=east] {{\tiny 0}};
\draw (1) node[anchor=east] {{\tiny 1}};
\draw (2) node[anchor=south] {{\tiny 2}};
\draw (3) node[anchor=south,color=white] {{\tiny 3}};
\draw (4) node[anchor=west,color=white] {{\tiny 4}};
\draw (5) node[anchor=west,color=white] {{\tiny 5}};
\end{tikzpicture}};

%2
\draw (4,0) node[rectangle,draw ]{
\begin{tikzpicture}[scale=0.4]
\coordinate (1) at (0,0);
\coordinate (0) at (0,2);
\coordinate (2) at (2,1);
\coordinate (3) at (4,1);
\coordinate (5) at (6,0);
\coordinate (4) at (6,2);
% the unions
\draw[fill=white] (0)--(1)--(2)--cycle;
% The points
\draw[fill=black] (0) circle (1pt and 1pt);
\draw[fill=black] (1) circle (1pt and 1pt);
\draw[fill=black] (2) circle (1pt and 1pt);
\draw[fill=white,color=white] (3) circle (1pt and 1pt);
\draw[fill=white,color=white] (4) circle (1pt and 1pt);
\draw[fill=white,color=white] (5) circle (1pt and 1pt);
% The names of the nodes
\draw (0) node[anchor=east] {{\tiny 0}};
\draw (1) node[anchor=east] {{\tiny 1}};
\draw (2) node[anchor=south] {{\tiny 2}};
\draw (3) node[anchor=south,color=white] {{\tiny 3}};
\draw (4) node[anchor=west,color=white] {{\tiny 4}};
\draw (5) node[anchor=west,color=white] {{\tiny 5}};
\end{tikzpicture}};

\draw (8,0) node[rectangle,draw ]{
\begin{tikzpicture}[scale=0.4]
\coordinate (1) at (0,0);
\coordinate (0) at (0,2);
\coordinate (2) at (2,1);
\coordinate (3) at (4,1);
\coordinate (5) at (6,0);
\coordinate (4) at (6,2);
% The unions among the points
\draw[fill=white] (0)--(1)--(2)--cycle;
% The points
\draw[fill=black] (0) circle (1pt and 1pt);
\draw[fill=black] (1) circle (1pt and 1pt);
\draw[fill=black] (2) circle (1pt and 1pt);
\draw[fill=black] (3) circle (1pt and 1pt);
\draw[fill=black] (4) circle (1pt and 1pt);
\draw[fill=black] (5) circle (1pt and 1pt);
% The names of the nodes
\draw (0) node[anchor=east] {{\tiny 0}};
\draw (1) node[anchor=east] {{\tiny 1}};
\draw (2) node[anchor=south] {{\tiny 2}};
\draw (3) node[anchor=south] {{\tiny 3}};
\draw (4) node[anchor=west] {{\tiny 4}};
\draw (5) node[anchor=west] {{\tiny 5}};
\end{tikzpicture}};

\draw (0,-2) node[rectangle,draw ]{
\begin{tikzpicture}[scale=0.4]
\coordinate (1) at (0,0);
\coordinate (0) at (0,2);
\coordinate (2) at (2,1);
\coordinate (3) at (4,1);
\coordinate (5) at (6,0);
\coordinate (4) at (6,2);
% The unions among the points
\draw[fill=white] (0)--(1)--(2)--cycle;
\draw[fill=white] (3)--(4)--(5)--cycle;
% The points
\draw[fill=black] (0) circle (1pt and 1pt);
\draw[fill=black] (1) circle (1pt and 1pt);
\draw[fill=black] (2) circle (1pt and 1pt);
\draw[fill=black] (3) circle (1pt and 1pt);
\draw[fill=black] (4) circle (1pt and 1pt);
\draw[fill=black] (5) circle (1pt and 1pt);
% The names of the nodes
\draw (0) node[anchor=east] {{\tiny 0}};
\draw (1) node[anchor=east] {{\tiny 1}};
\draw (2) node[anchor=south] {{\tiny 2}};
\draw (3) node[anchor=south] {{\tiny 3}};
\draw (4) node[anchor=west] {{\tiny 4}};
\draw (5) node[anchor=west] {{\tiny 5}};
\end{tikzpicture}};

\draw (4,-2) node[rectangle,draw ]{
\begin{tikzpicture}[scale=0.4]
\coordinate (1) at (0,0);
\coordinate (0) at (0,2);
\coordinate (2) at (2,1);
\coordinate (3) at (4,1);
\coordinate (5) at (6,0);
\coordinate (4) at (6,2);
% The unions among the points
\draw[fill=white] (0)--(1)--(2)--cycle;
\draw (2)--(3);
\draw[fill=white] (3)--(4)--(5)--cycle;
% The points
\draw[fill=black] (0) circle (1pt and 1pt);
\draw[fill=black] (1) circle (1pt and 1pt);
\draw[fill=black] (2) circle (1pt and 1pt);
\draw[fill=black] (3) circle (1pt and 1pt);
\draw[fill=black] (4) circle (1pt and 1pt);
\draw[fill=black] (5) circle (1pt and 1pt);
% The names of the nodes
\draw (0) node[anchor=east] {{\tiny 0}};
\draw (1) node[anchor=east] {{\tiny 1}};
\draw (2) node[anchor=south] {{\tiny 2}};
\draw (3) node[anchor=south] {{\tiny 3}};
\draw (4) node[anchor=west] {{\tiny 4}};
\draw (5) node[anchor=west] {{\tiny 5}};
\end{tikzpicture}};

\draw (8,-2) node[rectangle,draw ]{
\begin{tikzpicture}[scale=0.4]
\coordinate (1) at (0,0);
\coordinate (0) at (0,2);
\coordinate (2) at (2,1);
\coordinate (3) at (4,1);
\coordinate (5) at (6,0);
\coordinate (4) at (6,2);
% The unions among the points
\draw[fill=gray!80!white] (0)--(1)--(2)--cycle;
\draw (2)--(3);
\draw[fill=white] (3)--(4)--(5)--cycle;
% The points
\draw[fill=black] (0) circle (1pt and 1pt);
\draw[fill=black] (1) circle (1pt and 1pt);
\draw[fill=black] (2) circle (1pt and 1pt);
\draw[fill=black] (3) circle (1pt and 1pt);
\draw[fill=black] (4) circle (1pt and 1pt);
\draw[fill=black] (5) circle (1pt and 1pt);
% The names of the nodes
\draw (0) node[anchor=east] {{\tiny 0}};
\draw (1) node[anchor=east] {{\tiny 1}};
\draw (2) node[anchor=south] {{\tiny 2}};
\draw (3) node[anchor=south] {{\tiny 3}};
\draw (4) node[anchor=west] {{\tiny 4}};
\draw (5) node[anchor=west] {{\tiny 5}};
\end{tikzpicture}};
\end{tikzpicture}};
\end{tikzpicture}
\caption{Filtration of the diabolo simplicial complex}\label{filtration}
\end{figure}
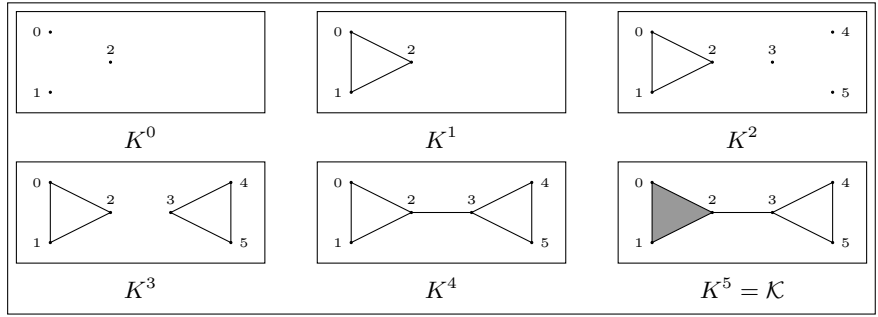

An example of a filtration can be seen in Figure~\ref{filtration} taking the diabolo complex of
Figure~\ref{diabolo} as ${\mathcal K}$.

Given a filtration of a simplicial complex and the $j$-th component of the filtration, let us say $K^j$,
we will denote $C_n(K^j), Z_n(K^j)$ and $B_n(K^j)$ by $C_n^j,Z_n^j$ and $B_n^j$ respectively. Therefore, we can represent
the chain complexes associated with a filtration using the following diagram.

$$\xymatrix{
\vdots \ar[d]_{d_3^0} &\vdots \ar[d]_{d_3^1}&\vdots \ar[d]_{d_3^2}& \\
C_2^0 \ar@{^(->}[r]^{i^0_2}\ar[d]_{d_2^0} & C_2^1 \ar@{^(->}[r]^{i_2^1}\ar[d]_{d_2^1} & C_2^2 \ar@{^(->}[r]^{i_2^2}\ar[d]_{d_2^2} & \ldots \\
C_1^0 \ar@{^(->}[r]^{i^0_1}\ar[d]_{d_1^0}  & C_1^1\ar@{^(->}[r]^{i_1^1}\ar[d]_{d_1^1}  & C_1^2 \ar@{^(->}[r]^{i_1^2}\ar[d]_{d_1^2}  & \ldots \\
C_0^0 \ar@{^(->}[r]^{i^0_0} & C_0^1 \ar@{^(->}[r]^{i^1_0} & C_0^2 \ar@{^(->}[r]^{i_0^2} & \ldots }$$

\noindent where $i_n^j$ is the map induced by the inclusion between the $n$-simplices of $K^j$ and the ones of $K^{j+1}$. Moreover, for $j<p$
we will denote $i_n^{j,p}$ to the map induced by the inclusion between the $n$-simplices of $K^j$ and the ones of $K^p$. Now, we can introduce
the notion of persistent homology groups.

\begin{Defn}
The \emph{$p$-persistent $n$-th homology group} of $K^j$, denoted by $H_n^{j,p}$, is defined as the quotient
 $H_n^{j,p}=i_n^{j,p}(Z_n^j)/(B_n^{p}\cap i_n^{j,p}(Z_n^j))$.

The \emph{$p$-persistent $n$-th Betti number} of $K^j$, denoted by $\beta_n^{j,p}$, is defined as the \emph{rank} of $H_n^{j,p}$.
\end{Defn}

The elements of $H_n^{j,p}$ are the $n$-dimensional homology classes of $K^j$ which are still alive at $K^p$. Hence
$\beta_n^{j,p}$ measures the number of $n$-dimensional classes of $K^j$ which are still alive at $K^p$. If we are
interested in computing the $n$-dimensional homology classes which \emph{are born} at $K^j$ and \emph{die} entering $K^p$, we have
the following formula:

\begin{equation}\label{eq1}
 \mu_n^{j,p} = (\beta_n^{j,p-1} - \beta_n^{j,p}) - (\beta_n^{j-1,p-1} - \beta_n^{j-1,p}).
\end{equation}

\noindent The first difference on the right hand side of the above formula measures the number of $n$-dimensional classes which are born
at or before $K^j$ and die entering $K^p$, and the second one counts the number of $n$-dimensional classes which are born
at or before $K^{j-1}$ and die entering $K^p$. Now, we can state the Fundamental Lemma of Persistent Homology.

\begin{Thm}[Fundamental Lemma of Persistent Homology]\label{thm:flph1}
Let $K^0 \subseteq K^1 \subseteq \ldots \subseteq K^m = \mathcal{K}$ be a filtration. For every pair of
indices $0\leq k \leq l \leq m$ and every dimension $n$, the $l$-persistent $n$-th Betti number of $K^k$ is
$$\beta_n^{k,l} = \sum_{0\leq i \leq k} \sum_{l<j\leq m} \mu_n^{i,j} + \beta_n^{k,m}.$$
\end{Thm}

This version of the Fundamental Lemma of Persistent Homology is stated slightly different from the one
presented in~\cite[pp.152]{EH10}; however both of them are equivalent. There are two differences between
the statement of Edelsbrunner and Harer~\cite[pp.152]{EH10} and
the one presented in Theorem~\ref{thm:flph1}. The former is related to the definition of $\mu_n^{j,p}$ which is defined
for the case $p\in \mathbb{N}$ and has to be extended for $p = \infty$. In particular for such a case,
we have the following formula:

$$ \mu_n^{j,\infty} = \beta_n^{j,\infty} - \beta_n^{j-1,\infty} $$

\noindent where $\beta_n^{j,\infty}$ is defined as the $n$-dimensional classes which are born at or before $K^j$ and
never die (or die in the $\infty$). It is worth noting that $\beta_n^{j,\infty}$ is equal to~$\beta_n^{j,m}$
(where $m$ is the last level of the filtration).

The latter difference is a consequence of the former one and is related to the inner sum of the theorem which
will be an infinite sum. In particular, the Edelsbrunner and Harer's Fundamental Lemma of Persistent Homology
is stated as follows.

\begin{Thm}[Fundamental Lemma of Persistent Homology]~\cite[pp.152]{EH10}\label{thm:flph}
Let $K^0 \subseteq K^1 \subseteq \ldots \subseteq K^m = \mathcal{K}$ be a filtration. For every pair of
indices $0\leq k \leq l \leq m$ and every dimension $n$, the $l$-persistent $n$-th Betti number of $K^k$ is
$$\beta_n^{k,l} = \sum_{0\leq i \leq k} \sum_{l<j} \mu_n^{i,j}.$$
\end{Thm}

As we have said previously, both formulations of the theorems are equivalent; however, the one presented
in Theorem~\ref{thm:flph1} is more suitable to be formalized since we do not need to handle infinite sums in
the \Coq/\ssr~theorem prover.

\begin{figure}
\centering
\begin{tikzpicture}

\draw (-2.5,2) node {$\beta_0$};

\draw (0,0) node{
\begin{tikzpicture}[scale=.5]
\draw[step=1,gray!20!white,very thin,fill=white] (0,0) grid (7,7);
\draw[-latex] (0,0) -- (0,7);
\draw[-latex] (0,0) -- (7,0);
\draw \foreach \s in {0,...,5}{(1+\s,-0.3) node {{\tiny $K^\s$}}};
\draw \foreach \s in {0,...,5}{(-0.3,1+\s) node {{\tiny $(\s)$}}};
\draw[*-o] (3,6) -- (4,6);
\draw[*-o] (3,5) -- (4,5);
\draw[*-o] (3,4) -- (5,4);
\draw[*-o] (1,3) -- (2,3);
\draw[*-o] (1,2) -- (2,2);
\draw[*->] (1,1) -- (6,1);
\end{tikzpicture}};

\draw (3.5,2) node {$\beta_1$};

\draw (6,0) node{
\begin{tikzpicture}[scale=.5]
\draw[step=1,gray!20!white,very thin,fill=white] (0,0) grid (7,3);
\draw[-latex] (0,0) -- (0,3);
\draw[-latex] (0,0) -- (7,0);
\draw \foreach \s in {0,...,5}{(1+\s,-0.3) node {{\tiny $K^\s$}}};

\draw (-1.5,1) node {{\tiny $(0)-(1)-(2)$}};
\draw (-1.5,2) node {{\tiny $(3) - (4) - (5)$}};

\draw[*->] (4,2) -- (6,2);
\draw[*-o] (2,1) -- (6,1);
\end{tikzpicture}};
\end{tikzpicture}
\caption{Barcodes of the diabolo filtration}\label{barcode}
\end{figure}
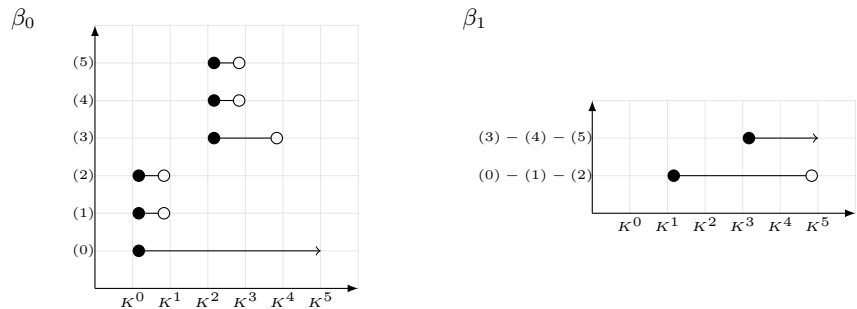

Finally, in order to shed light on the meaning of persistent homology, we introduce the usual way of
visualizing persistence. The lifetime of a $n$-dimensional homology class can be represented as an interval; namely,
a homology class which is born at level $K^i$ of the filtration and dies entering $K^j$ (with $i<j$) is
represented as the interval $[i,j)$, and if it is born at level $K^i$ but never dies we use the interval $[i,\infty)$.
A \emph{barcode} is defined to be the set of resulting intervals of a filtration. For example, the barcodes
in degree $0$ and $1$ associated with the filtration of Figure~\ref{filtration} are the ones depicted
in Figure~\ref{barcode}.

The barcodes in Figure~\ref{barcode} should be read as follows: In the case of the $\beta_0$ barcode,
the vertex $(0)$ is a connected component which \emph{is born} at level $0$ of the filtration and lives
forever; on the contrary, for example, the vertex $(3)$ \emph{is born} at level $2$ of the filtrations and
\emph{dies} entering the level $4$ when it is merged with another connected component. In the case of the $\beta_1$ barcode
(where holes are the $1$-homology classes), a hole appears at level $1$ of the filtration between the edges $(0,1), (0,2), (1,2)$ 
and dies entering the last
level of the filtration when it is filled with the triangle $(0,1,2)$; on the other hand, the hole which appears at
level $3$ of the filtration never dies.

%% file: abstract_algorithm.tex
Let us now introduce an abstract formalization of the notions presented in the previous section using
\Coq~together with the \ssr~extension.

\subsection{Simplicial Complexes and Homology}

In previous work, see~\cite{Formath11-ismfics,HDMMPS12}, we have formalized the notions presented
in subsections~\ref{math:sc} and~\ref{math:cc}. However, for the sake of clarity of the exposition we
include the main definitions and results which have been developed previously.

We begin with the notions related to simplicial complexes. The set of vertices
is represented by a finite type \lstinline?V?. A simplex is defined as a finite set of vertices.
Using this, the definition of a simplicial complex as a set of simplices closed under inclusion is straightforward:

\begin{lstlisting}
Variable V : finType.
Definition simplex := {set V}.
Definition simplicial_complex (c : {set simplex}) :=
  forall x, x \in c -> forall y : simplex, y \subset x -> y \in c.
\end{lstlisting}

The definition of the n-th incidence matrix of a simplicial complex, which is called \lstinline?incidence_mx_n?, takes two
arguments: a set of simplices \lstinline?c? and the dimension \lstinline?n?, and returns a {\sc SSReflect} matrix.
Moreover, we have proved the \emph{nilpotency condition} (see Definition~\ref{def:cc}) for two consecutive
incidence matrices encoded with \lstinline?incidence_mx_n?.

\begin{lstlisting}
Theorem incidence_matrices_sc_product:
  forall (V:finType) (n:nat) (sc: {set (simplex V)}),
    simplicial_complex sc ->
      (incidence_mx_n sc n) *m (incidence_mx_n sc (n.+1)) = 0.
\end{lstlisting}

The notion of homology is defined in {\sc Coq} as follows. Let $F$ be a field,
$V1, V2, V3$ vector spaces on $F$, and $f : V1 \rightarrow V2$, $g : V2 \rightarrow V3$
linear applications such that $g\circ f = 0$; then, the Homology of $f$ and $g$ is the quotient
between the kernel of $g$ and the image of $f$. This can be defined in {\sc Coq} in the
following way.

\begin{lstlisting}
Variables (F : fieldType) (V1 V2 V3 : vectType F)
           (f : linearApp V1 V2) (g : linearApp V2 V3).
Definition Homology := ((lker g) :\: (limg f)).
Definition Betti := \dim Homology.
\end{lstlisting}

Finally, this definition of homology can be instantiated for the homology in degree $n$ of a simplicial complex $sc$
using the linear applications associated with the incidence matrices in dimension $n+1$ and $n$ (given a matrix
\lstinline?M?, \lstinline?linearApp(M)? builds the linear application associated with \lstinline?M?).

\begin{lstlisting}
Definition Homology_sc_n (sc : {set simplex}) (n : nat) :=
  Homology (LinearApp (incidence_mx_n c n.+1))
            (LinearApp (incidence_mx_n c n)).
\end{lstlisting}

Analogously, we can define the $n$-th Betti number of a simplicial complex $sc$ instantiating the \lstinline?Betti?
definition.

\begin{lstlisting}
Definition Betti_sc_n (sc : {set simplex}) (n : nat) :=
  Betti (LinearApp (incidence_mx_n c n.+1))
         (LinearApp (incidence_mx_n c n)).
\end{lstlisting}

\subsection{Persistent homology}

In this section, we formalize the results presented in Subsection~\ref{math:ph}. First of all, we
define a more generic notion than the one of the persistent homology group $H_n^{j,p}$ associated
with a filtration. Such a notion involves the elements presented in the following diagram.

$$\xymatrix{
  & V_3 \ar[d]_{g} \\
V_1 \ar[r]^{i}\ar[d]_{f}  & V_4   \\
V_2 &  }$$

\noindent where $V_1$, $V_2$, $V_3$ and $V_4$ are vector spaces over a field $F$, $f : V_1 \rightarrow V_2$ and $g:V_3 \rightarrow V_4$ are
linear applications and $i:V_1 \rightarrow V_3$ is an injective linear
application. Using this we can define the vector space $P_{f,g,i}$ as the
quotient $i(\ker(f))/(im(g) \cap i(\ker(f)))$. We use the \lstinline?vector?
library of {\sc SSReflect} \cite{vector} in order to define this notion.

\begin{lstlisting}
Variable (F : fieldType).

Variables (V1 V2 V3 V4 : vectType F)
           (f : linearApp V1 V2) (g : linearApp V3 V4)
           (i : linearApp V1 V4).

Hypothesis (i_inj : injective i).

Definition P_fgi :=
   (i @: (lker f)) :\: ((limg g) :&: (i @: (lker f))).
\end{lstlisting}

As $i$ is an injective linear application and $(im(g) \cap i(\ker(f)))$ is a subspace of $i(\ker(f))$, the dimension
of $P_{f,g,i}$ is equal to the dimension of $\ker(f)$ (which in turn is equal to the dimension of $V_1$ minus the
dimension of $im(f)$) minus the dimension of $im(g) \cap i(\ker(f))$. This definition and its correctness
are introduced in {\sc Coq} as follows:

\begin{lstlisting}
Definition PBetti :=
   vdim V1 - \dim (limg f) - \dim ((limg g) :&: (i @: (lker f))).

Lemma PBettiE : \dim P_fgi = PBetti.
\end{lstlisting}

We omit the formal correctness proof for readability, we refer the interested
reader to look at the actual formalization~\cite{HCMS12}.

Let us now present how we instantiate these definitions for the persistent
homology group $H_n^{j,p}$ associated with a filtration. A filtration is defined
as a sequence of sets of simplices satisfying both that every element of the
sequence is a simplicial complex and that the elements of the sequence are
sorted using the subset relationship.

\begin{lstlisting}
Definition filtration (f : seq {set simplex V}) :=
    (forall x, x \in f -> simplicial_complex x) /\
    (forall i j, i <= j -> i < size f -> j < size f ->
         (nth set0 f i) \subset (nth set0 f j))
\end{lstlisting}

In order to define the inclusion matrix $i_n^{j,p}$ of a filtration, we first introduce the more generic
notion of inclusion matrix of two finite sets of simplices.

Representing a matrix requires an indexing of the simplices in \lstinline?Left? (for the
rows) and \lstinline?Top? (for the columns). Since \lstinline?Left? and \lstinline?Top? are finite sets,
they are equipped with a canonical enumeration: \lstinline?(enum_val Left i)? returns the \lstinline?i?-th
element of the set \lstinline?Left?. A coefficient $a_{ij}$ of the inclusion matrix will be 1 if the
\lstinline?i?-th simplex of \lstinline?Left? is equal to the \lstinline?j?-th simplex of \lstinline?Top? and
0 otherwise.

Therefore, we can define the inclusion matrix of two finite sets of simplices as follows.

\begin{lstlisting}
Variables Left Top : {set simplex}.

Definition inclusionMatrix :=
  \matrix_(i < #|Left|, j < #|Top|)
    if (enum_val i == enum_val j) then 1 else 0:'F_2.
\end{lstlisting}

In the definition above, it can be noted that the first argument of \lstinline?enum_val?
is implicit since it is determined by the context. Indeed, the notation \lstinline?i < #|Left|?
means that the type of \lstinline?i? is \lstinline?'I_(#|Left|)?, that is \lstinline?i?
is an ordinal ranging from $0$ to \lstinline?#|Left|-1?, where \lstinline?#|X|? denotes
the cardinality of the set \lstinline?X?. With this type information, the system expands
\lstinline?enum_val i? to \lstinline?enum_val Left i?, thus resolving the ambiguity
(and similarly for \lstinline?j?).

The type annotation \lstinline?0:'F_2? indicates that the 0 and 1 appearing as coefficients
of the matrix are the two elements of \lstinline?F_2?, that is, $\mathbb{Z}_2$ as a field.
We now define the inclusion matrix $i_n^{j,p}$ of a filtration \lstinline?f? by
instantiating \lstinline?Left? and \lstinline?Top? to the set of $n$-simplices of the \lstinline?j? and, respectively,
\lstinline?p? component of \lstinline?f?.

\begin{lstlisting}
Variable f : (seq {set (simplex V)}).
Variables n j p : nat.

Definition n_simplices (c : {set (simplex V)}):=
  [set x \in c | #|x|==n.+1].
Definition n_k_simplices (k : nat) :=
  n_simplices (nth set0 f k) n.
Definition inclusion_matrix_n_j_p :=
  inclusionMatrix (n_k_simplices j) (n_k_simplices p).
\end{lstlisting}

The main result that we have proved about \lstinline?inclusion_matrix_n_j_p? is that the linear application
associated with it is injective.

\begin{lstlisting}
Lemma injective_LinearApp_inclusion_mx :
  injective (LinearApp (inclusion_matrix_n_j_p)).
\end{lstlisting}

Now we have all of the necessary ingredients to define the persistent homology group $H_n^{j,p}$ and the persistent
Betti number $\beta_n^{j,p}$ just instantiating \lstinline?P_fgi? and \lstinline?PBetti? with the linear applications
associated with the matrices $d_n^j$, $d_{n+1}^{p}$ and $i_n^{j,p}$. These matrices are encoded as
\lstinline?(incidence_mx_n (nth set0 f j)? \lstinline?n)?, \lstinline?(incidence_mx_n (nth set0 f p)? \lstinline?n.+1)? and
\lstinline?(inclusion_mx f n j p)? respectively.

\begin{lstlisting}
Variable (V:finType) (f: (seq {set (simplex V)})) (n j p:nat).
Hypothesis f_is_filtration : filtration f.
Hypothesis j_is_in_filtration :  j < size f.
Hypothesis j_leq_p_is_in_filtration : j <= p < size f.

Definition p_persistent_n_homology_K_j :=
 P_fgi (LinearApp (incidence_mx_n (nth set0 f j) n)))
        (LinearApp (incidence_mx_n (nth set0 f p) n.+1))
        (LinearApp (inclusion_mx f n j p)).

Definition p_persistent_n_betti_K_j :=
 PBetti (LinearApp (incidence_mx_n (nth set0 f j) n))
         (LinearApp (incidence_mx_n (nth set0 f p) n.+1))
         (LinearApp (inclusion_mx f n j p)).
\end{lstlisting}

Finally, we can define $\mu_n^{j,p}$ (see Formula~\ref{eq1}) and prove our version of
the Fundamental Lemma of Persistent Homology (Theorem~\ref{thm:flph1}).

\begin{lstlisting}
Theorem fundamental_lemma_persistent_homology (k l : nat) :
  \sum_(0<=j<k.+1) (\sum_(l.+1 <= p < (size f).+1) (mu n j p)) =
  (p_persistent_n_betti_K_j f n j p) -
  (p_persistent_n_betti_K_j f n j (size f)).
\end{lstlisting}

The \lstinline?bigop? library of SSReflect, see~\cite{BGOP08}, has played a key role in the proof of the above theorem. This library
is devoted to generic indexed big operations, like $\sum\limits_{i=0}^n f(i)$ or $\bigcap\limits_{i\in I} f(i)$, and
their properties. Again, the interested reader can consult the whole development of the formal proof of the Fundamental Lemma
of Persistent Homology in~\cite{HCMS12}.

%% file: effective.tex
One of the goals of this work was the development of certified programs to
compute both Betti and persistent Betti numbers. In the previous section we have
provided the definitions of such notions given in terms of linear applications of
vector spaces. However, we do not usually work with linear applications when computing
Betti and persistent Betti numbers but with the matrices representing those linear
applications.

Equation~\ref{betti_rank} provides the explicit formula to compute Betti numbers
from two matrices. So we can use it to define this new notion using \ssr~matrices
(where~\lstinline?'M[K]_(m,n)? is a $m \times n$ matrix over \lstinline?K?) and
prove that this new notion is equivalent to the one given by the \lstinline?Betti?
definition.

\begin{lstlisting}
Definition Betti_rank (mxf:'M[K]_(v1,v2)) (mxg:'M[K]_(v2,v3)) :=
  v2 - \rank mxg - \rank mxf.

Lemma Betti_rankE  (v1 v2 v3 : nat) (mxf:'M[K]_(v1,v2))
  (mxg:'M[K]_(v2,v3)), mxf *m mxg = 0 ->
  dim_homology mxf mxg = Betti (LinearApp mxf) (LinearApp mxg).
\end{lstlisting}

Similarly, we can define persistent Betti numbers in terms of matrices and prove the equivalence
between such a definition and the one given in \lstinline?PBetti? definition.

\begin{lstlisting}
Definition PBetti_rank (mxf:'M[K]_(v1,v2)) (mxg:'M[K]_(v3,v4))
  (mxi:'M[K]_(v1,v4)) :=
   (v1 - \rank mxf - (\rank mxg + (v1 - \rank mxf) -
   \rank (col_mx mxg (kermx mxf) *m mxi)))%N.

Lemma PBetti_rankE (v1 v2 v3 v4 :nat) : forall (mxf:'M[K]_(v1,v2))
 (mxg:'M[K]_(v3,v4)) (mxi:'M[K]_(v1,v4)),
  injective (LinearApp mxi) ->
   PBetti_rank =
   PBetti (LinearApp mxf) (LinearApp mxg) (LinearApp mxi).
\end{lstlisting}

However the use of \ssr~libraries may trigger heavy computations during deduction steps, that would not
terminate within a reasonable amount of time. To handle this issue some definitions, like matrices, are locked in
a way that do not allow direct computations.

To overcome this pitfall, we use the methodology presented in~\cite{DMS12} whose key idea is the one of
\emph{refinements}. Roughly speaking, the correctness of mathematical algorithms are proved using all the
high-level theory available in the \ssr~libraries and then the algorithms are refined to an implementation
on simpler data structures that will be the ones running on the machine. In our particular case of matrices we use lists of lists as the low level data type for representing them.

The methodology presented in~\cite{DMS12} have been implemented as a new library, built on top of \ssr~libraries,
which is called CoqEAL~\cite{COQEAL_CODE}. This library includes the refinements of almost all the
algorithms involved in the computation of Betti and persistent Betti numbers. To be more concrete, the only
algorithm which has been necessary to refine is the one in charge of computing the row kernel of a matrix.

The \lstinline?kermx? function is already available in the \ssr~library
and implements the row kernel of a matrix. This algorithm has been refined into an efficient version, called \lstinline?ker?,
which works with abstract matrices. The equivalence between both algorithms has been proved in the following lemma.

\begin{lstlisting}
Lemma eqmx_ker m n (M : 'M[K]_(m,n)) : ker M :=: kermx M.
\end{lstlisting}

\noindent where the notation \lstinline?A :=: B? means that \lstinline?A? and \lstinline?B? are equivalent matrices. Subsequently,
the \lstinline?ker? algorithm has been translated to the low level data type, list of lists, with the function
\lstinline?ker_seqmx? and, eventually, we have ensure that \lstinline?ker_seqmx? perform the same operation
that its high-level counterpart \lstinline?ker?.

Now, we have all the necessary programs to define an executable version of both Betti and persistent Betti numbers
and prove the equivalence with their high-level versions.

\begin{lstlisting}
Definition ex_Betti_rank (v1 v2 v3:nat) (mxf mxg:seqmatrix K) :=
  v2 - (rank_elim_seqmx v2 v3 mxg) - (rank_elim_seqmx v1 v2 mxf).

Definition ex_PBetti_rank (v1 v2 v3 v4 : nat)
 (mxf mxg mxi : seqmatrix K) :=
  let rf := rank_elim_seqmx v1 v2 mxf in
  let rg := rank_elim_seqmx v3 v4 mxg in
  v1 - rf - (rg + (v1 - rf) -
  rank_elim_seqmx (v3 + size (ker_seqmx v1 v2 mxf)) v4
       (col_seqmx mxg (mulseqmx (ker_seqmx v1 v2 mxf) mxi))).

Lemma ex_Betti_rankE: forall (v1 v2 v3 : nat) (mxf: 'M[K]_(v1,v2)) (mxg : 'M[K]_(v2,v3)),
   ex_Betti_rank v1 v2 v3 (seqmx_of_mx K mxf) (seqmx_of_mx K mxg) = Betti_rank mxf mxg.

Lemma ex_PBetti_rank_PBetti_rank_E : forall (v1 v2 v3 v4 :nat)
 (mxf: 'M[K]_(v1,v2)) (mxg : 'M[K]_(v3,v4)) (mxi : 'M[K]_(v1,v4)),
   ex_PBetti_rank v1 v2 v3 v4 (seqmx_of_mx K mxf) (seqmx_of_mx K mxg) (seqmx_of_mx K mxi) = PBetti_rank mxf mxg mxi.
\end{lstlisting}

\noindent It is worth noting that the executable functions on matrices, represented as lists of lists, usually need the
size of the matrices, as can be seen for instance in the \lstinline?rank_elim_seqmx? function. Moreover, the function
\lstinline?seqmx_of_mx? is the one in charge of transforming abstract matrices into lists of lists (which are encoded
as the type \lstinline?seqmatrix?).

Following the same pattern, we have defined executable simplicial complexes and their connection with the computation
of Betti and persistent Betti numbers. In particular, the \lstinline?ex_Betti_sc? function takes as argument a simplicial
complex \lstinline?c? and a natural number \lstinline?n? and computes the \lstinline?n?-th Betti number of \lstinline?c?
and the \lstinline?ex_p_persistent_n_Betti_K_j? function takes as argument a filtration \lstinline?f? and three natural
numbers \lstinline?p,n,j? and computes the \lstinline?p? persistent \lstinline?n?-th Betti number of the \lstinline?j? level
of the filtration \lstinline?f?. Some examples of the usage of these functions are introduced in the following section.

%% file: experimental.tex
In this section we try to clarify how Betti and persistent Betti numbers can be computed within {\sc Coq}.
Let us start with the computation of Betti numbers of the simplicial complex of Figure~\ref{diabolo}.

Simplicial complexes are built in {\sc Coq} providing their \emph{facets}. A \emph{facet} of a simplicial
complex ${\mathcal K}$ is a maximal simplex with respect to the subset order $\subseteq$ among the simplices
of ${\mathcal K}$. To construct the simplicial complex associated with a sequence of facets,
$\mathcal{F}$, we generate all the faces of the simplices of $\mathcal{F}$; subsequently, if we perform the
set union of all the faces we obtain the simplicial complex associated with $\mathcal{F}$. This procedure
have been implemented, and its correctness have been proved, using {\sc Coq} in~\cite{Formath11-ismfics}.
In the case of the diabolo complex of Figure~\ref{diabolo} its facets are: $\{(2,3), (3,4), (3,5), (4,5), (0,1,2)\}$.

The procedure to compute Betti numbers of the diabolo complex is as follows. First, we
define the list of facets:

\begin{lstlisting}
Definition diabolo_facets := [::[::2;3];[::3;4];[::3;5];[::4;5];[::0;1;2]].
\end{lstlisting}

\noindent and, subsequently, we compute Betti numbers (of dimension $0$ and $1$) using the
instruction \lstinline?ex_Betti_sc? which takes as arguments the facets of the simplicial complex
and the dimension.

\begin{lstlisting}
Eval vm_compute in ex_Betti_sc diabolo_facets 0.
Eval vm_compute in ex_Betti_sc diabolo_facets 1.
\end{lstlisting}

\noindent obtaining in both cases 1 (this means that the diabolo complex has a connected component
and a hole) in just milliseconds.

The procedure to compute persistent Betti numbers is quite similar. First of
all, we define the filtration providing the facets of each one of the simplicial
complexes of the filtration. In the case of the diabolo filtration of
Figure~\ref{filtration}, the representation in {\sc Coq} of the filtration is
the following one:

\begin{lstlisting}
Definition diabolo_filtration :=
   [::[::[::0];[::1];[::2]];
   [::[::0;1];[::1;2];[::0;2]];
   [::[::0;1];[::1;2];[::0;2];[::3];[::4];[::5]];
   [::[::0;1];[::1;2];[::0;2];[::3;4];[::4;5];[::3;5]];
   [::[::0;1];[::1;2];[::0;2];[::3;4];[::4;5];[::3;5];[::2;3]];
   [::[::2;3];[::3;4];[::3;5];[::4;5];[::0;1;2]]].
\end{lstlisting}

Using this, we can compute persistent Betti numbers by calling the function \lstinline?ex_p_persistent_n_betti_K_j?.
Therefore, we can combine the functions to compute Betti and persistent Betti numbers in order to obtain
information about the filtration. For instance, if we want to know how many connected components which live at the
level $0$ of the filtration (this is computed by \lstinline?(nth nil diabolo_filtration 0)?) are still alive at
level $4$, we use the following instructions:

\begin{lstlisting}
Eval vm_compute in ex_Betti_sc (nth nil diabolo_filtration 0) 0.
Eval vm_compute in
  ex_p_persistent_n_Betti_K_j diabolo_filtration 0 0 4.
\end{lstlisting}

\noindent obtaining as result $3$ and $1$ respectively. This means that there are three connected components
at level $0$ of the filtration but just one of them is still alive at level 4.

As a benchmark to test the efficiency of {\sc Coq} programs, we have considered several random simplicial
complexes and filtrations generated from a fixed number of triangles. The results can be seen in Table~\ref{table}
where we show for each number of triangles the times (in seconds) to compute both Betti and persistent Betti numbers.

\begin{table}
{%
\begin{center}
\begin{tabular}{|l|c|c|c|c|c|}
\hline
 & 10 & 50 & 100 & 200 & 500\\
\hline
Betti & 0.024 & 2 & 18 & 146 & 1856\\
\hline
Persistent Betti & 0.036 & 3 & 25 & 190 & 2731\\
\hline
\end{tabular}
\end{center}
}%
\caption{Execution times}\label{table}
\end{table}

Of course, our {\sc Coq} programs take much more time to compute Betti and persistent Betti numbers
than special purpose software packages such as Chomp~\cite{Chomp} and the GAP homology package~\cite{GAP-homology}
for Betti numbers or \emph{JavaPlex}~\cite{javaPlex} and Dionysus~\cite{Dionysus} for persistent Betti numbers. However, it is
worth remarking that {\sc Coq} is an Interative Theorem Prover and in this kind of systems, unlike Computer Algebra
systems or special purpose packages, efficient computational capabilities have not been the main goal up to now. %% implementation of efficient algorithms was put in the background
%% up to now.

Nevertheless, things are changing and there is an on-going effort in the implementation of efficient mathematical
algorithms running inside {\sc Coq}. In this line, we can highlight the works on machine integers and
arrays~\cite{armand_extending_2010}, efficient real numbers~\cite{gregoire_compiled_2002} or an approach which
consists in internally compiling {\sc Coq} terms to the functional programming language
{\sc OCaml}~\cite{mathieu_boespflug_full_2011}.

%% file: conclusions.tex
In this paper we have presented a set of formally verified programs which allows us to \emph{effectively} compute
persistent Betti numbers within {\sc Coq}. To carry out this task, it has been necessary a formalization of the
basic notions related to persistent homology. Moreover, we have
formalized relevant theorems like the Fundamental Lemma of Persistent Homology. This illustrates that
Interactive Theorem Provers are mature enough to tackle the formalization of theories
in non-trivial mathematical domains. This fact can be seen also in the proof of the
Four Color Theorem~\cite{FCT}, in the Flyspeck project~\cite{flyspeck}, devoted to
the formal proof of the Kepler conjecture~\cite{Hales05}; or in the classification of
finite groups~\cite{MathComp}.

One of our main concerns for the future is associated with the formalization of efficient
mathematical algorithms. This is a necessary effort which has to be carried out before
undertaking other of our goals: the application of our programs to biomedical problems.

Homological techniques have been successfully applied in the biomedical context,
see~\cite{vessel,Mrozek12,HDMMPS12}. In this
environment it is necessary to have both efficient and reliable software systems; therefore the
use of formally verified efficient algorithms seems desirable.

It is also appealing to use this work as a basis for further developments. We can tackle
the formalization of different extensions of persistent homology; for instance,
\emph{multidimensional persistence}~\cite{Carlsson2007} or \emph{ZigZag persistence}~\cite{Zigzag}.

In summary we can conclude that we are working towards an efficient formal library of Computational Algebraic Topology,
in this line of work we can mention the formalizations of Effective Homology~\cite{ABR08,DR2011} and
Discrete Morse Theory~\cite{HPR12}, however more work is still necessary to reach our goal.

%% file: cphwcs.bbl
\begin{thebibliography}{10}

\bibitem{Chomp}
Chomp: Computational homology project.
\newblock Software available at \url{http://chomp.rutgers.edu/software/}.

\bibitem{MathComp}
{Mathematical components team homepage}.
\newblock \url{http://www.msr-inria.inria.fr/Projects/math-components}.

\bibitem{ABR08}
J.~Aransay, C.~Ballarin, and J.~Rubio.
\newblock A mechanized proof of the {B}asic {P}erturbation {L}emma.
\newblock {\em Journal of Automated Reasoning}, 40(4):271--292, 2008.

\bibitem{armand_extending_2010}
M.~Armand, B.~Grégoire, A.~Spiwack, and L.~Théry.
\newblock {Extending Coq with Imperative Features and Its Application to {SAT}
  Verification}.
\newblock In {\em Proceedings Interactive Theorem Proving 2010 (ITP'2010)},
  volume 6172 of {\em Lecture Notes in Computer Science}, pages 83--98, 2010.

\bibitem{BGOP08}
Y.~Bertot, G.~Gonthier, S.~O. Biha, and I.~Pasca.
\newblock {Canonical Big Operators}.
\newblock In {\em Proceedings 21st International Conference on Theorem Proving
  in Higher Order Logics (TPHOLS'08)}, volume 5170 of {\em Lecture Notes in
  Computer Science}, pages 86--101, 2008.

\bibitem{Zigzag}
G.~Carlsson and V.~de~Silva.
\newblock Zigzag persistence.
\newblock {\em CoRR}, abs/0812.0197, 2008.

\bibitem{Carlsson2007}
G.~Carlsson and A.~Zomorodian.
\newblock The theory of multidimensional persistence.
\newblock In {\em Proceedings of the 23rd annual symposium on Computational
  geometry (SCG '07)}, pages 184--193. ACM, 2007.

\bibitem{Coq}
{\textsc{Coq} development team}.
\newblock {The \textsc{Coq} Proof Assistant, version 8.4}.
\newblock Technical report, 2012.

\bibitem{SG07}
V.~de~Silva and R.~Ghrist.
\newblock Homological sensor networks.
\newblock {\em Notices of the American Mathematical Society}, 54(1):10--17,
  2007.

\bibitem{DE95}
C.~J.~A. Delfinado and H.~Edelsbrunner.
\newblock {An incremental algorithm for Betti numbers of simplicial complexes
  on the 3 sphere}.
\newblock {\em Computer Aided Geometry Design}, 12:771--784, 1995.

\bibitem{DG06}
T.~K. Dey and S.~Goswami.
\newblock Provable surface reconstruction from noisy samples.
\newblock {\em Computational Geometry: Theory and Applications},
  35(1):124--141, 2006.

\bibitem{DR2011}
C.~Dom\'inguez and J.~Rubio.
\newblock {Effective Homology of Bicomplexes, formalized in Coq}.
\newblock {\em Theoretical Computer Science}, 412:962--970, 2011.

\bibitem{GAP-homology}
J.~Dumas, F.~Heckenbach, B.~D. Saunders, and V.~Welker.
\newblock Gap homology package.
\newblock Software available at \url{http://www.linalg.org/gap.html}, 2002.

\bibitem{COQEAL_CODE}
M.~Dénès, A.~M\"ortberg, and V.~Siles.
\newblock {CoqEAL, the Coq Effective Algebra Library}, 2012.
\newblock \url{http://www-sop.inria.fr/members/Maxime.Denes/coqeal}.

\bibitem{DMS12}
M.~Dénès, A.~Mörtberg, and V.~Siles.
\newblock {A Refinement Based Approach to Computational Algebra in Coq}.
\newblock In {\em Proceedings Interactive Theorem Proving 2012 (ITP'2012)},
  volume 7406 of {\em {Lectures Notes in Computer Science}}, pages 83--98,
  2012.

\bibitem{EH10}
H.~Edelsbrunner and J.~L. Harer.
\newblock {\em Computational Topology: An Introduction}.
\newblock American Mathematical Society, 2010.

\bibitem{ELZ02}
H.~Edelsbrunner, D.~Letscher, and A.~Zomorodian.
\newblock {Topological persistence and simplification}.
\newblock {\em Discrete Computional Geometry}, 28:511--533, 2002.

\bibitem{FL99}
P.~Frosini and C.~Landi.
\newblock {Syze theory as a topological tool for computer vision}.
\newblock {\em Pattern Recognition and Image Analysis}, 9:596--603, 1999.

\bibitem{barcode08}
R.~Ghrist.
\newblock Barcodes: the persistent topology of data.
\newblock {\em Bulletin American Mathematical Society}, 45:61--75, 2008.

\bibitem{FCT}
G.~Gonthier.
\newblock {\em Formal proof - The Four-Color Theorem}, volume~55.
\newblock Notices of the American Mathematical Society, 2008.

\bibitem{vector}
G.~Gonthier.
\newblock Point-free, set-free concrete linear algebra.
\newblock In {\em Proceedings Interactive Theorem Proving 2011 (ITP'2011)},
  volume 6898 of {\em Lecture Notes in Computer Science}, pages 103--118, 2011.

\bibitem{SSReflect}
G.~Gonthier and A.~Mahboubi.
\newblock {An introduction to small scale reflection in Coq}.
\newblock {\em Journal of Formal Reasoning}, 3(2):95--152, 2010.

\bibitem{gregoire_compiled_2002}
B.~Gr{\'e}goire and X.~Leroy.
\newblock A compiled implementation of strong reduction.
\newblock In {\em International Conference on Functional Programming 2002},
  pages 235--246. ACM Press, 2002.

\bibitem{Hales05}
T.~Hales.
\newblock {A proof of the Kepler conjecture}.
\newblock {\em Annals of Mathematics}, 162:1065--1185, 2005.

\bibitem{flyspeck}
T.~Hales.
\newblock The flyspeck project fact sheet.
\newblock Project description available at
  \url{http://code.google.com/p/flyspeck/}, 2005.

\bibitem{HCMS12}
J.~Heras, T.~Coquand, A.~M\"ortberg, and V.~Siles.
\newblock Formalization of homology and persistent homology, 2012.
\newblock
  \url{http://wiki.portal.chalmers.se/cse/pmwiki.php/ForMath/ProofExamples#wp3%
ex6}.

\bibitem{HDMMPS12}
J.~Heras, M.~D\'en\`es, G.~Mata, A.~M\"ortberg, M.~Poza, and V.~Siles.
\newblock {Towards a certified computation of homology groups for digital
  images}.
\newblock In {\em Proceedings 4th International Workshoph on Computational
  Topology in Image Context (CTIC'2012)}, volume 7309 of {\em Lecture Notes in
  Computer Science}, pages 49--57, 2012.

\bibitem{Formath11-ismfics}
J.~Heras, M.~Poza, M.~D\'en\`es, and L.~Rideau.
\newblock Incidence simplicial matrices formalized in {Coq/SSReflect}.
\newblock In {\em Proceedings 18th Symposium on the Integration of Symbolic
  Computation and Mechanised Reasoning (Calculemus'2011)}, volume 6824 of {\em
  Lecture Notes in Computer Science}, pages 30--44, 2011.

\bibitem{HPR12}
J.~Heras, M.~Poza, and J.~Rubio.
\newblock {Verifying an algorithm computing Discrete Vector Fields for digital
  imaging}.
\newblock In {\em Proceedings 19th Symposium on the Integration of Symbolic
  Computation and Mechanised Reasoning (Calculemus'2012)}, volume 7362 of {\em
  Lecture Notes in Computer Science}, pages 215--229, 2012.

\bibitem{Basic-Algebra}
N.~Jacobson.
\newblock {\em Basic Algebra II}.
\newblock W. H. Freeman and Company, 2nd edition, 1989.

\bibitem{cubicalhomology}
G.~Kedenburg.
\newblock Persistent cubical homology in pattern recognition.
\newblock Diplomarbeit. Universität Hamburg, 2010.

\bibitem{MacLane}
S.~MacLane.
\newblock {\em Homology}.
\newblock Springer, 1963.

\bibitem{mathieu_boespflug_full_2011}
{{Mathieu} Boespflug}, {{Maxime} Dénès}, and {{Benjamin} Grégoire}.
\newblock {Full Reduction at Full Throttle}.
\newblock In {\em Proceedings Certified Programs and Proofs}, volume 7086 of
  {\em Lecture Notes in Computer Science}, pages 362--377, 2011.

\bibitem{Dionysus}
D.~Morozov.
\newblock Dionysus.
\newblock Software available at \url{http://www.mrzv.org/software/dionysus/},
  2012.

\bibitem{Mrozek12}
M.~Mrozek et~al.
\newblock Homological methods for extraction and analysis of linear features in
  multidimensional images.
\newblock {\em Pattern Recognition}, 45(1):285--298, 2012.

\bibitem{vessel}
M.~Niethammer et~al.
\newblock Analysis of blood vessel topology by cubical homology.
\newblock {\em Image Rochester NY}, 2(2):969--972, 2002.

\bibitem{R99}
V.~Robins.
\newblock {Towards computing homology from finite approximations}.
\newblock {\em Topology proceedings}, 24:503--532, 1999.

\bibitem{javaPlex}
A.~Tausz, M.~Vejdemo-Johansson, and H.~Adams.
\newblock Javaplex: A research software package for persistent (co)homology.
\newblock Software available at \url{http://code.google.com/javaplex}, 2011.

\bibitem{ZC05}
A.~Zomorodian and G.~Carlsson.
\newblock {Computing Persistent Homology}.
\newblock {\em Discrete and Computional Geometry}, 33:249--274, 2005.

\end{thebibliography}
